\documentclass[a4paper,10pt]{article}
\usepackage[english]{babel}
\usepackage[utf8]{inputenc}

\usepackage{authblk}
\usepackage[margin=1.2in]{geometry}
\usepackage{graphicx} 
\graphicspath{{figures/}} 
\usepackage{sidecap}
\usepackage{amsmath}
  
\author[a,b,1,2]{George Datseris}
\author[c,1]{Annika Ziereis}
\author[c]{Thorsten Albrecht} 
\author[c]{York Hagmayer} 
\author[a, b, d]{Viola Priesemann} 
\author[a,b, d]{Theo Geisel}
\affil[a]{Max Planck Institute for Dynamics and Self-Organization, 37077 Göttingen, Germany}
\affil[b]{Department of Physics, Georg-August-University Göttingen, 37073 Göttingen, Germany}
\affil[c]{Georg-Elias-Mueller Institute for Psychology, Georg-August-University Göttingen, 37073 Göttingen, Germany}
\affil[d]{Bernstein Center for Computational Neuroscience, 37077 Göttingen, Germany}
\affil[1]{G.D. and A.Z. contributed equally to this work}
\affil[2]{george.datseris@ds.mpg.de}

\title{Does it Swing? \\Microtiming Deviations and Swing Feeling in Jazz}
\begin{document}

\maketitle

\begin{abstract}
Jazz music that swings has the fascinating power to elicit a pleasant sensation of flow in listeners and the desire to synchronize body movements with the music. Whether microtiming deviations (MTDs), i.e. small timing deviations below the bar or phrase level, enhance the feeling of swing is highly debated in the current literature. Studies on other groove related genres did not find evidence for a positive impact of MTDs. The present study addresses jazz music and swing in particular, as there is some evidence that microtiming patterns are genre-specific. We recorded twelve piano jazz standards played by a professional pianist and manipulated the natural MTDs of the recordings in systematic ways by quantizing, expanding and inverting them. MTDs were defined with respect to a grid determined by the average swing ratio. The original and manipulated versions were presented in an online survey and evaluated by 160 listeners with various musical skill levels and backgrounds. Across pieces the quantized versions (without MTDs) were rated slightly higher and versions with expanded MTDs were rated lower with regard to swing than the original recordings. Unexpectedly, inversion had no impact on swing ratings except for two pieces. Our results suggest that naturally fluctuating MTDs are not an essential factor for the feeling of swing.
\end{abstract}

\section{Introduction}

The swing feeling is a major feature in most jazz music performances, but what is it that actually makes a piece of music swing? For many years, musicologists have attempted to describe and explain this phenomenon.
Already in 1937, the rhythmic characteristic of swing was described as "an almost imperceptibly hurried accent of the second and fourth beats" and further, ``the word really defies definition, for most musicians protest that swing cannot be regarded wholly as a matter of rhythm, but that it depends rather on a personal emotional response to rhythm that cannot be written in any musical notation.''~\cite{Nye1937} 
The swing feeling is sometimes classified as a kind of groove in jazz music \cite{Levitin2018, Proegler1995}. Swing and groove are not generally interchangeable (there is no swing without groove, but there may be groove without swing).  Nevertheless  both concepts are related. Groove is commonly defined as the musical aspect that induces a pleasant sensation (enjoyment) of wanting to move along with the music (entrainment or flow) \cite{Janata2012}. 

Rhythmic characteristics of a piece, like syncopation~\cite{Sioros2014, Witek2014} or the swing ratio~\cite{Friberg2002,Butterfield2011,Wesolowski2016,Benadon2009a}, are known to influence groove.
Specifically in jazz the \emph{swing ratio}, i.e. the length ratio of consecutive 8\textsuperscript{th} notes in the typical long-short pattern~\cite{Friberg2002}, has been studied extensively~\cite{Butterfield2011}.
The variation of the swing ratio ratio within a piece and between musicians has been related to musical variables like intervals, articulation, harmony and melody \cite{Wesolowski2016}. It is also conjectured to emphasize important structural aspects of the music, to draw attention to a soloist's phrase \cite{Benadon2009a}, and to facilitate the perception in multiple onsets of notes by contrasting timbres.

The influence of the swing ratio, along with other minute timing deviations known as \emph{microtiming deviations} (MTDs), on the swing feeling and groove has generated considerable research interest~\cite{Madison2011, Butterfield2010, Madison2014, Fruehauf2013, Davies2013, Senn2016, Matsushita2016, Hofmann2017, Kilchenmann2015,Benadon2009, Butterfield2011, Danielsen2015, Danielsen2015a, Senn2017, McGuiness2006, Iyer2002}.
Although professional musicians can play extremely accurately, they will not perfectly match a hypothetical systematic grid produced by a metronome. 
Many musicians accept that MTDs are important for the swing feeling as reflected in Charles Keil's theory of \textit{participatory discrepancies} (PD-Theory,\cite{Keil1987}). According to this theory it "is the little discrepancies within a jazz drummer's beat, between bass and drums, between rhythm section and soloists, that create `swing' and invite us to participate" \cite{Keil1987}. From this perspective, MTDs create tension \cite{Keil1966, Proegler1995}, liven up the music and elicit swing, rather than reflecting inaccuracies.

On the other hand, empirical studies performed on funk, samba, jazz or rock music have failed to find a positive impact of MTDs on groove \cite{Madison2011, Butterfield2010, Madison2014, Fruehauf2013, Davies2013}. Music without MTDs was rated more ``groovy'' than music with MTDs~\cite{Fruehauf2013, Davies2013} and increasing the size of MTDs consistently led to worse groove ratings~\cite{Davies2013, Senn2016, Matsushita2016, Hofmann2017, Kilchenmann2015}. In only two studies, patterns with very small MTDs resulted in higher groove ratings than no MTDs~\cite{Senn2016, Hofmann2017}. The exact pattern of MTDs was shown to be impactful~\cite{Fruehauf2013, Matsushita2016, Senn2017}, but empirical research so far did not show a positive impact of MTDs on the groove feeling. Attempts to explain origins of this relation are based on predictive timings~\cite{Merker2009}, the complexity of the music structure~\cite{Fitch2007, Grahn2007, Chmiel2017}, and syncopation~\cite{Song2013, Sioros2014, Witek2014}. It remains unclear, however, whether these results directly apply to MTDs in swinging music.

Besides in empirical and qualitative studies, musical MTDs have been studied quantitatively by analyzing naturally occurring MTDs ~\cite{Hennig2011, Hennig2014, Raesaenen2015, Sogorski2018}. Specifically in~\cite{Hennig2011} it was shown that the sequence of human MTDs from a metronome tick is not random, but instead power-law correlated. This finding applies not only to simple tapping-like time series, but also to real music recordings~\cite{Raesaenen2015, Sogorski2018} and musicians' interactions~\cite{Hennig2014}. Interestingly, the variability of MTDs in jazz music was found to be larger than in rock~\cite{Sogorski2018}, which may be a hint of MTDs having larger impact in jazz music and for the swing feeling.

\begin{figure}
    \centering
    \includegraphics[width=\textwidth]{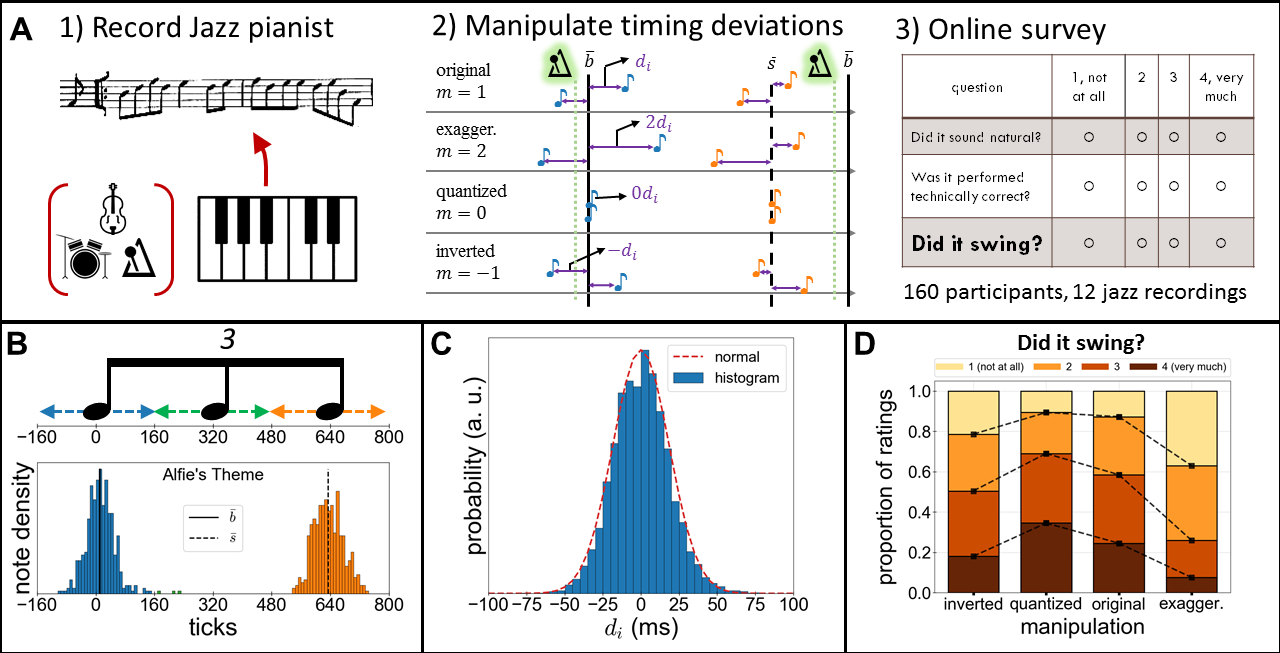}
    \caption{\textbf{A}1) Experimental setup. A professional pianist was recorded performing jazz standards, while listening to quantized bass and drum tracks. 2) We determined the average base note position $\bar{b}$ and swing note position $\bar{s}$ for each recording. The microtiming deviations $d_i$ of the notes were defined with respect to $\bar{b}, \bar{s}$ and were subsequently manipulated using three different manipulations: exaggerated, quantized and inverted. 3) Original and manipulated recordings were then used in an online survey where musicians judged them. \textbf{B} (top) Sketch of the range of ticks that an 8\textsuperscript{th} note triplet covers. (bottom) Note position density (modulo the quarter note), and mean note positions $\bar{b}, \bar{s}$. Note classification: blue=base ($b$), orange=swing ($s$), green=disregarded (for details see sec.~\ref{sec:MTDs_def_2}). \textbf{C} Histogram of micro-timing deviations, measured in milliseconds, across all pieces and both note types $b, s$. Plotted is also a normal distribution with the same mean (0.08 ms) and variance (18.39 ms). \textbf{D} A result of the online survey: proportion of answers averaged across pieces and participants (dashed lines are guide to the eye reflecting cumulative proportions of answers).}
    \label{fig:main}
\end{figure}

\subsection{Aim and Research Question}
There is some evidence that both the listeners' familiarity and expectation of microtiming patterns as well as the MTD patterns vary across genres and that jazz music, compared to other groove related music, elicits different expectations with regard to (micro-)timing patterns \cite{Senn2017, Davies2013, Senn2016, Fruehauf2013, Sogorski2018}. Nevertheless the role of MTDs for the swing feeling has remained quite controversial as outlined above. In the present paper we therefore address the question, whether naturally occurring MTDs of a soloist do enhance the swing feeling. This is done by carrying out an online survey for which we manipulated the MTDs of original performances. In line with the (widespread) opinion that musicians ``can feel it, but you just can't explain it''~\cite[p.8-10]{Treadwell} we let 160 listeners, e.g. professional and semiprofessional musicians, rate the swing feeling in different versions of manipulated and original recordings. We chose a broad variety of twelve jazz pieces (see Table~\ref{tab:MTDs}), differing in several aspects like swing ratio, tempo and syncopation. This ensures that our findings generalize across different pieces and do not depend on specific aspects. Besides quantizing, i.e. eliminating the MTDs, and augmenting them by a factor of 2, we also inverted them (by multiplying by a factor of -1) thereby changing the respective microtiming pattern without changing its magnitude. The process we just described is also shown in Fig.~\ref{fig:main}\textbf{A}.

In our listening experiment we applied a mixed within-between-subjects design in which only one version for each piece was presented. We thus obtained a global assessment of the swing feeling of the individual pieces and avoided prompting our participants (to try) to detect differences between manipulations. We aimed for a diverse sample of jazz and non-jazz, expert and non-expert listeners (see \ref{sec:sample}) to control for possible genre- and skill-dependent moderation effects. A more detailed description of the listening study is presented in the Supplementary Information. 
According to PD theory, the original versions of the pieces should swing the most, having the optimal MTDs pattern. In contrast, empirical research suggests that the quantized versions elicit the strongest groove due to higher predictability of the pulse. From both perspectives we expect a reduced swing feeling for versions with exaggerated MTDs. Furthermore, if there was an objectively ``appropriate'' use of MTDs to create swing, inversion should lead to lower swing ratings as the pianist's specific style of playing would be erased. 

\subsection{Defining Micro-Timing Deviations} \label{sec:def_MTDs}

A detailed definition of the MTDs, based on the internal format of MIDI data, is presented in the methods sec.~\ref{sec:MTDs_def_2}. In summary, the pianist was recorded effectively with a metronome (see methods sec.~\ref{sec:recordings}). As the recorded pieces were in a Jazz style, the phrasing of notes is based on 8\textsuperscript{th} note triplets, out of which only the first and third notes are played (``swung 8\textsuperscript{th} notes''). This is also confirmed by the note density of a recording shown in Fig.~\ref{fig:main}~\textbf{B}. 
As the swing ratio is variable and not always the ``typical'' value of 2:1 (i.e. perfect triplets, see also Table~\ref{tab:MTDs} and methods sec.~\ref{sec:MTDs_def_2}), there is an ambiguity in defining the MTDs. We found the following definition based on an average swing ratio to be most appropriate for this study.

Let us first consider the first notes of all triplets, which we call ``base'' notes $b$. These notes are supposed to follow the metronome click, but of course are not played exactly on it and instead deviate (a sketch is also shown in Fig.~\ref{fig:main}). However, it may be that the pianist is intentionally playing behind or ahead of the click, and we would want to keep this aspect of the performance intact. What we did was calculate the average base note position $\bar{b}$ with respect to the metronome click individually for each recording. Then, the MTDs of the base notes $d_i$ is simply their deviation from their mean position $\bar{b}$ (modulo the quarter note). This is illustrated in Fig.~\ref{fig:main}\textbf{A} (middle panel, top row, blue notes) and \textbf{B} (blue notes, first part of triplet). The same process was applied to the third notes of all triplets, which we call ``swing'' notes $s$. 
For them, the MTDs are measured with respect to $\bar{s}$, see Fig.~\ref{fig:main} (orange colors).
An advantage of defining the MTDs this way is that it respects the \emph{average} swing ratio of the recording (see results section).

After establishing the MTDs, we manipulated them by multiplying each deviation $d_i$ with a constant factor $m$. 
In this study we chose $m = 0, 2, -1$. This process is shown in Figure~\ref{fig:main}\textbf{A} middle panel.
Notes that fell in between the first and third part of an 8\textsuperscript{th} note triplet were not manipulated. However, they represented on average only 0.7\% of the total notes and at most 1.8\% (Serenade to a Cuckoo). Using a manipulation of $m=0$ closely corresponds to the so-called ``quantization'' procedure, a well-known process in music production. Our ``quantization'', however, keeps intentional playing consistently ahead or behind the beat intact (if there is any). Lastly, we point out that no other aspect of the performance was changed. The pianist's choice of what chords to play, what phrasing to use, how to anticipate notes, etc. was left untouched. 
In addition, note intensity (dynamic level, e.g. piano or forte), duration and even sustain pedal usage were left intact.

\section{Results}

\begin{table}
\centering 
	\begin{tabular}{l|c|c|c|c|c|c|c|c|c} 
		Recording               & BPM & $\bar{b}$ & $\bar{s}$ & $\sigma_b$ (ms) & $\sigma_s$ (ms)& $r$  & $\delta_r$ & $\rho_{b\to s}$  & $\rho_{s\to b}$   \\ \hline
        Alfie's Theme        & 135  & 11      & 634     & 16.94  & 18.3   & 1.99 & 0.39       & 0.22 & 0.51 \\
        Blue Monk            & 140  & 0       & 627     & 18.85  & 17.5   & 1.93 & 0.39       & 0.43 & 0.57 \\
        Don't Get Around*    & 140  & 24      & 644     & 20.44  & 19.18  & 2.11 & 0.51       & 0.48 & 0.51 \\
        Doxy                 & 130  & 1       & 624     & 15.31  & 17.85  & 1.89 & 0.33       & 0.21 & 0.59 \\
        Four                 & 170  & 3       & 602     & 17.7   & 17.01  & 1.73 & 0.38       & 0.67 & 0.75 \\
        In a Mellow Tone     & 160  & -7      & 623     & 20.1   & 21.79  & 1.94 & 0.58       & 0.72 & 0.69 \\
        Jordu                & 150  & -1      & 603     & 19.11  & 17.31  & 1.73 & 0.33       & 0.39 & 0.66 \\
        Now's The Time       & 190  & -20     & 582     & 15.55  & 21.01  & 1.62 & 0.45       & 0.53 & 0.73 \\
        Paper Moon           & 135  & -2      & 635     & 18.73  & 17.34  & 2.0  & 0.41       & 0.41 & 0.46 \\
        Serenade to a Cuckoo & 140  & 22      & 634     & 21.6   & 20.15  & 2.0  & 0.42       & 0.1  & 0.59 \\
        So What              & 160  & 4       & 582     & 18.48  & 15.59  & 1.57 & 0.32       & 0.67 & 0.8  \\
        Yardbird Suite       & 180  & 7       & 596     & 21.12  & 20.38  & 1.71 & 0.51       & 0.59 & 0.53

	\end{tabular} 
	\caption{Properties of micro-timing deviations (MTDs) of recordings. Notice that mean note positions $\bar{b}, \bar{s}$ are given in ticks, but the standard deviations of the MTDs $\sigma_b, \sigma_s$ are in milliseconds (see~\ref{sec:MTDs_def_2}). $r$ and $\delta_r$ are the average swing ratio and its standard deviation, respectively (see~\ref{sec:swing_ratio}).
	$\rho_{\alpha\to \beta}$ is Spearman's rank correlation coefficient between microtiming deviations of pairs of notes of types $\alpha, \beta$ (see~\ref{sec:correlations}). All recording versions used are available as Supplemental Material. *Title of piece ``Don't Get Around Much Anymore'' was shortened to fit.} 
	\label{tab:MTDs} 
\end{table} 

\subsection{Statistical properties of the recordings} First, we investigated the basic statistical properties of MTDs (Table~\ref{tab:MTDs}).
Across pieces, the MTDs were approximately following a normal distribution (Figure~\ref{fig:main}\textbf{C}) with $\sigma \approx 18.5 \pm 2$ ms, for either $b, s$ across all pieces. The fairly invariant magnitude of the MTDs
is noteworthy, because the individual pieces differ in tempo, complexity and many other aspects. This suggests that the magnitude of these MTDs is probably not controlled by the performer, but reflects instead a ``human error'' from a temporally exact performance. Similar observations were made in previous studies \cite{Hennig2011,Sogorski2018}. 
In Fig.~\ref{fig:main} we show the computed distribution of MTDs of all played notes (all recordings and both $b, s$, in order to increase the amount of data). The result is compared with a normal distribution with same mean and std., showing an excellent fit. 

The average swing ratio $r$ characterizes the average position of a swing note $s$ with respect to the metronome click (see sec.~\ref{sec:swing_ratio}). In jazz music $r$ typically varies between 1:1 (straight 8\textsuperscript{th}s) to 3:1 (dotted-8\textsuperscript{th} and 16\textsuperscript{th} notes in alternation), depending on tempo and musician \cite{Friberg2002}. As expected, the swing ratio decreased as the tempo of the recordings increased (Table~\ref{tab:MTDs}). 
In our recordings, the maximum $|\bar{b}|$ is approximately 10 ms (24 ticks), 
which puts it at the limits of auditory thresholds. This means that playing ``behind'' or ``ahead of'' the beat was not pronounced in our recordings.

\begin{figure}
    \includegraphics[width=\columnwidth]{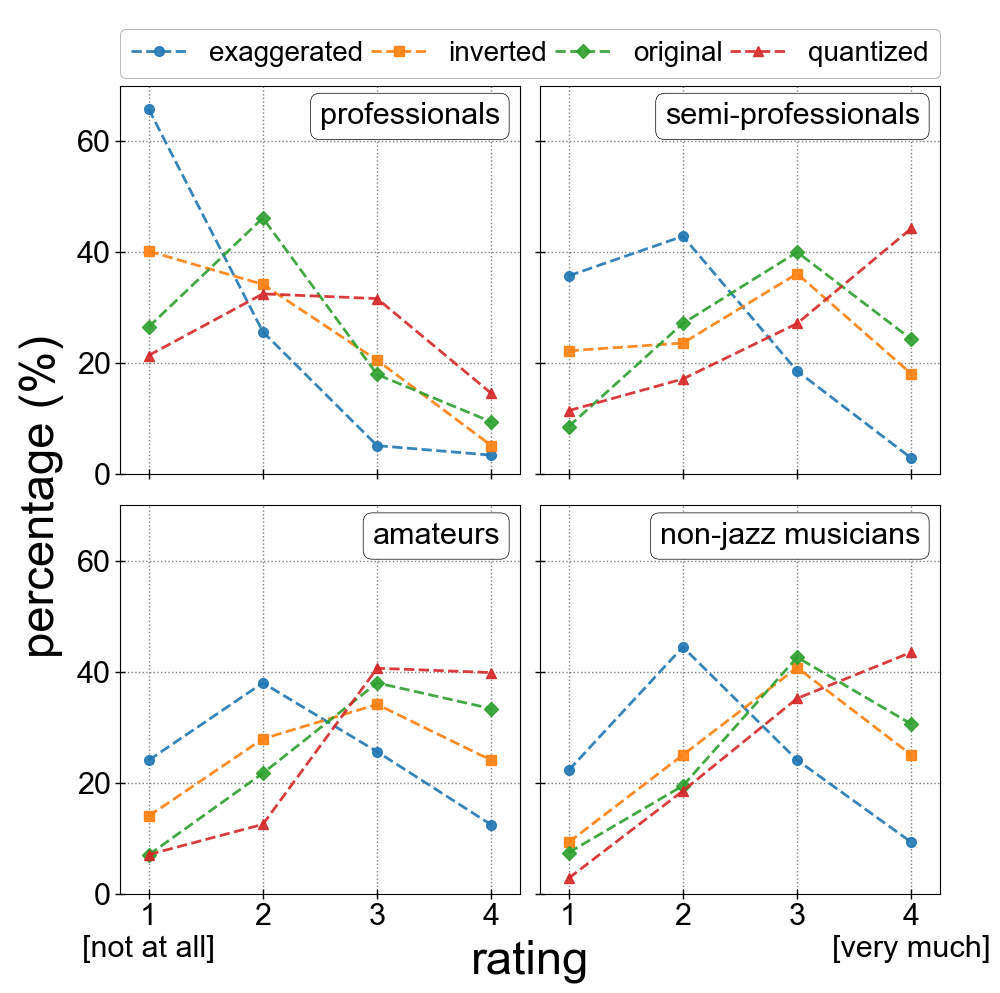}
    \caption{Percentages of ratings for the question \emph{``Does it swing?''} for the different manipulations, averaged across pieces. Swing ratings for the four different versions are displayed separately for the groups of musicians.}
    \label{fig:percentages}
\end{figure}

\subsection{Online survey} 

\paragraph{Participants.}
\label{sec:sample}
 In the online survey we had 160 participants (42 females, 115 males, 3 NA, \textit{M}$_{age}$ = 39.6 $\pm$ 15.5 years), who rated at least half of the pieces. Of those participants, 24.4\% were professional jazz, 15\% semi-professional jazz, 29.4\% amateur jazz and 26.3\% non-jazz musicians as well as 5\% jazz-loving non-musician. Non-musicians without a penchant for jazz were excluded from the survey.
 
\paragraph{Ratings.} The results of the online survey indicate that the manipulation of MTDs had an impact on swing ratings. Fig.~\ref{fig:percentages} displays the proportions of the participants' ratings on each point of the swing scale. Across all musicians' groups the exaggerated versions elicited the lowest swing ratings, as the peak of the distributions is shifted to the lower end of the swing rating scale. The opposite was found for the quantized versions, for which the majority of participants gave the highest swing ratings. The distribution for the inverted version is similar to the original. Both were rated as more swinging compared to the exaggerated but not as much as the quantized versions. Our statistical analysis (see below) did not find a significant difference between the ratings of the inverted and original versions. Additionally to the manipulation, a difference between the musicians' groups was found. Professional jazz musicians gave overall lower swing ratings, independent of the version of a piece. This can be seen in Fig.~\ref{fig:percentages}, where the distributions for professionals have more weight in the lower ratings in contrast to the other groups.

The detailed statistical results were obtained with a proportional odds mixed model that was fitted to the data (see Section \ref{sec:statanal}). The manipulation of MTDs (Likehood Ratio Tests: LRT(3) = 308.44 $p < .001$) and the musical background of participants (LRT(3) = 62.58 $p < .001$) were significant predictors of swing ratings.
Quantization and exaggeration of the MTDs led to significantly different swing ratings compared to the original recordings. The Odds Ratio (OR) for the quantized (compared to the original) version to be swinging were 1.65 (95\% CI = 1.03 - 2.65, $p = .037$) and for the exaggerated versions 0.23 (95\% CI = 0.14 - 0.36, $p <.001$), respectively. Hence, by averaging and removing the variability of MTDs, higher swing ratings were found, whereas by increasing the magnitude of MTDs the swing feeling was reduced. Averaging across all pieces, inversion of MTDs hardly changed swing ratings (\textit{OR} = 0.75, 95\% CI = 0.47 - 1.19, $p  = .224$). This non-significant effect of MTDs inversion is due to the majority of pieces, for which swing ratings were similar to those of the original recordings. For two out of twelve pieces (Jordu and Yardbird Suite), however, inversion had a negative impact on swing ratings.

Receiver Operating Characteristics (ROC) analysis were performed to test the discriminability of conditions. The effect of inversion on Jordu and Yardbird Suite was shown by areas under the curves (AUC) differing significantly from 0.5 (Jordu: \textit{AUC} = .32, 95\% CI = .20 - .44; Yardbird Suite: \textit{AUC} = .34, 95\% CI = .22 - .46). All ROC analyses can be obtained from the supplementary information. We also performed a sanity check, searching for possible correlations of the AUC of quantized versus original (which is a number indicating preference of quantized over original versions), with other aspects of the recordings, e.g. tempo, intensity, note density, etc.. Since we found no correlations, we present this information in the Supplemental Material.

With regard to naturalness, only the exaggerated version differed from the original versions and were perceived as less natural (\textit{OR}= 0.27; 95\% CI = 0.17 - 0.43, $p <.001$). 
Naturalness ratings of quantized (\textit{OR} = 0.98; 95\% CI = 0.61 - 1.57, $p = .935$) and inverted (\textit{OR} = 0.64; 95\% CI = 0.40 - 1.02, $p = .060$) versions were similar to the original version. 
Quantized and original versions were similarly rated as being played technically correct (\textit{OR} = 1.27; 95\% CI = 0.80 - 2.04, $p = .351$). Exaggerated (\textit{OR} = 0.08; 95\% CI = 0.05 - 0.13, $p <.001$) and inverted (\textit{OR} = 0.49; 95\% CI = 0.31 - 0.79, $p = .001$) versions were rated lower on this dimension. 

The musical background of participants showed a main effect. As a reference the group of non-jazz musicians was chosen. Professional jazz musicians were observed to be generally more harsh in their judgments and gave overall lower swing ratings (\textit{OR} = 0.16, 95\% CI = 0.08 - 0.32, $p < .001$). However, there is no interaction of musicians' group and condition (LRT(9)= 7.16 $p = .621$). Thus, the manipulation had similar effects independent of the participants' musical background. 

\section{Discussion}

In the present study we tested whether MTDs, originating from a professional musician performance, enhance the swing feeling. To do so, we manipulated the original recordings in a systematic way and compared the ratings of the manipulated and the original versions of the pieces. Our results show that MTDs have an impact on the swing feeling. Contrary to the PD-theory \cite{Keil1966}, however, naturally occurring MTDs do not seem essential for the swing feeling, as the versions without MTDs elicited the highest self-reported swing feeling. This is in line with other empirical studies, which suggest that a very tight and regular timing is more likely to enhance entrainment and groove \cite{Davies2013, Senn2016, Matsushita2016, Hofmann2017, Kilchenmann2015}. Our own study dealt with jazz music and the phenomenon of swing in particular. We imagine, however, that our conclusions might generalize also to the phenomenon of groove in various other music styles.

A common argument of musicians is that music without MTDs sounds unnatural and machine-like. Our approach of quantizing the original recordings tried to conserve as much as possible the style introduced by the pianist. We did not synchronize the pianist's track and the rhythm section to eliminate delay between the tracks. If the pianist chose to play consistently behind the beat to create a \textit{laid back feel}, our procedure maintained this asynchrony in the manipulations (even though this was not significant in our recordings, see $\bar{b}$ in Table~\ref{tab:MTDs}). Moreover, we conserved all other parts that one could regard as the ``human touch'' of the performance, like for example velocity and harmonization. The important difference between the original and quantized versions is that our quantization procedure eliminated fluctuations of the swing ratio within a piece and made it perfectly systematic and predictable from a temporal perspective. This might have facilitated the listeners' entrainment. To find out, whether our quantized versions would be perceived as less natural than the original recordings, we introduced a question to control for this. Our data did not show a difference between original and quantized versions regarding naturalness (Supplemental Information). Moreover, the two dimensions (swing and naturalness) were highly correlated (\textit{r} = .66) and should not be regarded as independent.

The present study showed that large deviations have a negative impact on swing, as the exaggerated versions with doubled MTDs obtained strongly decreased swing ratings. We know from previous research that the listeners' perceptual abilities and sensitivity for timing differences can moderate the impact of MTDs~\cite{Senn2016}. Consequently, professional jazz musicians, who may be assumed to be more sensitive to timing differences, were the ones with the strictest judgments (similar to Ref.~\cite{Davies2013}). Not only did they give lowest swing rating of all groups to the originals, but also to the quantized versions. We did not find qualitative differences in the effect of our manipulations, however, between musicians' groups. Like the other groups, professional jazz musicians rated the quantized versions as the most and the exaggerated versions as the least swinging.

\paragraph{Inverted MTDs}
Inverting the MTDs in a series of notes can have two different outcomes. One is that the \textit{inter-note intervals} (INI) are left unchanged because $d_i$ and $d_{i+1}$ have the same sign, for some $i$. The second possibility is that a pair of short-long intervals becomes long-short instead, a case of having $d_i$, $d_{i+2}$ with the same sign but $d_{i+1}$ with opposing sign.
For small magnitudes of $d_i$, the first case should not lead to perceptual differences. One might, however, expect that the second case can have significant impact on the rating of the listeners, as it would change the local swing ratio. In contrast to this expectation, our per-piece analysis showed that the effect was substantial only for 2 out of 12 pieces (see Supplemental Information) and in the other cases the ratings between inverted and original were very similar (still the original version was rated higher, on average). 

It is possible, however, to understand why original and inverted versions were rated similarly, if we consider that timing deviations $d_i$ are typically longe-range correlated and not random, as shown in  previous studies~\cite{Hennig2011, Raesaenen2015, Sogorski2018, Iyer2002}. This means that given some $d_i$, it is more probable that the subsequent $d_{i+1}, d_{i+2}$ are numerically similar, with the similarity (i.e. correlation function) decaying as a power-law as the distance from $i$ increases. 

In our case, to estimate correlations between subsequent $d_i$, we used Spearman's rank correlation coefficient $\rho$ (see section~\ref{sec:correlations} and Table~\ref{tab:MTDs}). Overall, correlations of the MTDs are high ($\rho \sim 0.5$), showing that it is much more probable for subsequent $d_i$ to have the same sign. Therefore, interchanging a short-long interval with a long-short only happened rarely, which explains the unexpectedly good rating of the inverted versions.

\section{Conclusions}
MTDs clearly exist in human playing, and are known to be power-law correlated~\cite{Hennig2011, Raesaenen2015, Sogorski2018}. It was highly debated, however, whether they have a positive impact on the perception of ``groove'', or ``swing''~\cite{Madison2011, Butterfield2010, Madison2014, Fruehauf2013, Davies2013, Senn2016, Matsushita2016, Hofmann2017, Kilchenmann2015,Benadon2009, Butterfield2011, Danielsen2015, Danielsen2015a, Senn2017, Madison2014, McGuiness2006, Iyer2002, Keil1987, Keil1966, Proegler1995}. By performing a study using a well controlled process of manipulating timing deviations (measured with respect to a grid determined by the mean swing ratio), as well as having a big sample pool, we are able to show that the MTDs as defined in this paper are not essential for the swing feeling. Instead they rather might have a negative impact.
We thus conclude that a rhythm should be persistent in its timing to yield a pronounced ``swing feeling''. 

We want to point out that we studied a laboratory/studio setting, where we recorded a pianist performing on top of a pre-recorded (and quantized) rhythm section. In live  performances in addition, musicians interact, synchronize, and adapt to each other. As we encouraged comments of the participants of our online survey, it was pointed out various times that these two aspects of interaction and adaptation were believed to be important (see supplementary information). Thus it still remains to be seen, whether our results also carry over to a setting of interacting musicians. We intend to investigate this point in future research. The situation studied in the present paper, where solo instruments are recorded on top of a pre-recorded rhythm section, nevertheless is very common in the music industry.

\section{Methods}

All raw piano MIDI files, audio samples used in the survey, the online survey results (as raw data) and the survey participant's comments on ``what defines swing'' are available online as Supporting Information. Computer code used for analyzing and manipulating the MIDI files is open source. We used the software MIDI.jl and MusicManipulations.jl which are hosted on GitHub and are MIT licensed, see~\cite{Datseris2019} for more details on the software.

\subsection{Recording and Manipulating}
\label{sec:recordings}

A unique professional bass and drum track was written for each piece, using human performances. Both were quantized to 8\textsuperscript{th} note triplets. The pianist was listening to these two tracks while being recorded, effectively playing with a metronome. In addition, the pianist had as much freedom as possible for the performance. The only restraint in the recordings was that ornaments (like e.g. trills, acciaccaturas, etc.) were to be avoided, as they don't have a well defined temporal position. It should be noted that the pianist \emph{could} play full 8\textsuperscript{th} note triplets, but he chose to play mostly swung 8\textsuperscript{th} notes.
Finally, we did not distinguish between notes played with the left hand (chords) or the right hand (melody), so they are all considered belonging to the same series of notes. 

For each recording the recorded MIDI piano track was exported and then manipulated as described in sec.~\ref{sec:def_MTDs}. It was then re-imported into our recording software (in this study Cubase) to produce audio files.

\subsection{Definition of Microtiming Deviations}
\label{sec:MTDs_def_2}
Since the recordings are in the form of MIDI data, we measure the temporal position of notes in \textit{ticks}.
This is a dimensionless unit of measuring time, native to music. Each tick is equal to $1/960$ of a quarter note and can be transformed to milliseconds once the beats per minute (BPM) are given (1 tick = (62.5/BPM) ms). As mentioned in the main text, we define MTDs individually for the ``base'' notes $b$ and ``swing'' notes $s$. Thus, as a first step, we classify note events:
Let $p_i$ be the true position of a note (in ticks) and
\begin{equation}
    f(p_i) = \begin{cases} 
   (p_i \,\text{mod}\, 960) - 960 & \text{if } (p_i \,\text{mod}\, 960) \ge 800 \\
   (p_i \,\text{mod}\, 960)     & \text{else}
  \end{cases}
  \label{eq:f}
\end{equation}
where $(a\, \text{mod}\, b)$ is the remainder after dividing $a$ by $b$. $f$ brings all notes into a single quarter note and its action is shown in Fig.~\ref{fig:main}\textbf{B}. 
Notes are then categorized according to which part of the 8\textsuperscript{th} note triplet they fall into: the \emph{set} of ``base'' notes $b$ has $f(b) \in [-160, 160)$, while the \emph{set} of ``swing'' notes $s$ has $f(s) \in [480, 800)$, see Fig.~\ref{fig:main}\textbf{B}.

In the second step, we calculate (for each note type) the mean note position as
\begin{equation}
    \bar{b} = \frac{1}{||b||}\sum_{i\in b} f(p_i)\,,\quad 
    \bar{s} = \frac{1}{||s||}\sum_{i\in s} f(p_i)\,
    \label{eq:meannoteposition}
\end{equation}
with $||\cdot||$ the cardinality of a set.
The MTDs of note $i$ is $d_i = f(p_i) - \bar{b}$ or $d_i = f(p_i) - \bar{s}$ depending on note category. Notes with $f(p_i) \in [160, 480)$ (i.e. they fall in the 2\textsuperscript{nd} part of the triplet) are so few that they are disregarded. To compute $\sigma$ in Table~\ref{tab:MTDs} we translate MTDs to milliseconds by multiplying them with (62.5/BPM) for each piece.

\subsection{Definition of the Average Swing Ratio}
\label{sec:swing_ratio}

The average swing ratio $r$ quantifies the ratio of the length of a base note to a swing note. However, because not all swing notes are preceded \emph{and} followed by a base note, we define each individual ratio as if base notes were played exactly on the metronome click. This can be justified since the mean base note position $\bar{b}$ is very close to 0 across all recordings, as can be seen from Table.~\ref{tab:MTDs}. The formula we used is
\begin{equation}
 r_i = \frac{(p_i \,\text{mod}\, 960)}{960 - (p_i \,\text{mod}\, 960)},\;i\in s.
 \label{eq:swingratio}
\end{equation}
$r$ is the average of all $r_i$ while $\delta_r$ is their standard deviation.

\subsection{Correlations of Microtiming Deviations}
\label{sec:correlations}
For all base notes $b$ we find pairs of base-swing notes, the distance of which is not greater than one quarter note. We then compute $\rho$ using all possible deviations $d_i$ and $d_{i+1}$ for all $i$ that satisfy the aforementioned criterion. The same process is done for pairs of swing notes followed by a base note, which are not necessarily identical. In the case where multiple notes are played in a given temporal bin (e.g for chords played together with a melody note), we always only use the timing information of the earliest note onset in the bin.

\subsection{Statistical Analyses of Online Survey}
\label{sec:statanal}
Because distances between points on the four point rating scale we used may not be equal, ordinal regression was used. Two random effects in the model allow for variation in judgments between participants and for variation between pieces. 
The cumulative link mixed model was fitted using the \textit{clmm} function from the ordinal package in \textit{R} \cite{Christensen2018}. The significance of the predictors was estimated by likelihood-ratio tests (LRT), which compared the likelihood of a full model with three fixed effects condition (original, quantized, exaggerated, inverted), musicians' category and their interaction against partial and no-covariate models in which one of the predictors was removed. The model can be written as
\begin{equation}
 \text{logit}\left(\frac{\text{P}(\text{score}(i, p) \leq j)}{1-\text{P}(\text{score}(i, p)\leq j)}\right) = \alpha_{j}+u_{ip} - \beta_{1}x -\beta_{2}y_{i} -\beta_{3}x * y_{i}
 \label{eq:linkmm}
\end{equation}
where $\text{P}(\text{score}(i, p) \leq j)$ gives the probability that the swing rating of participant $i$ for piece $p$ is $j$ or lower. Different $\beta_{1}$ are estimated for the three different manipulations (exaggerating, quantizing and inverting compared to the original version). Similarly, $\beta_{2}$ represents the estimated effect of participant $i$-th musicians' category. The $\alpha_{j}$ represents the cutpoints in the model and $u_{ip}$ are the random effects for participants and pieces. (Model parameters can be found in Supplemental Material)

The assumption of proportional odds was tested in two ways: (1) testing the models without random factor as cumulative against proportional odds models by vector generalized linear models (ref VGAM) and (2) visually, by plotting the differences between predicted logits for varying levels of the predictors for every outcome category (Supplemental Information). Both methods suggest excluding the group of jazz-loving non-musicians ($N=8$) from the analysis because including them results in a Hauck-Donner effect \cite{HauckJr1977} and variation in the proportional odds, which can lead to unstable estimates of the regression model.

\paragraph{Author contributions} Music recordings and first article draft: G.D. A.Z. 
Analysis \& manipulation of MTDs, production of audio: G.D. 
Design of the online study: A.Z., T.A., Y.H. 
Implementation and analysis of online study: A.Z. 
Conceiving the study: T.G., V.P.
All authors contributed to concept and design of the study, and the manuscript.
\paragraph{Competing interests} The authors declare no competing interests.
\paragraph{Acknowledgements} We would like to especially thank the pianist that performed the recordings, Uwe Meile, for the excellent performance, as well as drummer Tim Dudek for helpful discussions.

\bibliographystyle{unsrt}
\bibliography{references.bib}

\end{document}